\documentclass[manuscript]{aastex}

\begin{document}
\pagenumbering{arabic}

\title{WHAT ARE S0 (0) GALAXIES?}

\author{Sidney van den Bergh}
\affil{Dominion Astrophysical Observatory, Herzberg Institute of Astrophysics, National Research Council of Canada, 5071 West Saanich Road, Victoria, BC, V9E 2E7, Canada}
\email{sidney.vandenbergh@nrc-cnrc.gc.ca}

\begin{abstract}   

   Among early-type galaxies with almost circular
isophotes E0 and E1 galaxies are, at 99.3\% significance,
more luminous than face-on objects classified as S0 (0)
and S(0) (1). This result supports the view that
rotation and ``diskiness'' are more important in the
outer regions of faint early-type galaxies than they
are for more luminous galaxies of very early
morphological type.

\end{abstract}

\keywords{galaxies:  classification}

\section{INTRODUCTION}

    Hubble (1936, p.44) introduced the classification type
S0 to bridge the gap between objects of types E7 and Sa. 
Galaxies that are morphologically intermediate between ellipticals 
and spirals have also been referred to as ``lenticulars'' by de 
Vaucouleurs (1959). A detailed discussion of the morphology of S0 
galaxies is given in Sandage (1961, p.10) and in Sandage \& Bedke 
(1994, Vol.1, p.7). Not unexpectedly S0 galaxies, on average, appear 
more flattened than E galaxies. Following Hubble this flattening {\it f} may 
be defined as  {\it f} = 10({\it a-b})/{\it a}, where {\it a} and {\it b} are the major and minor
axis diameters, respectively. In the convention adopted by
Sandage \& Tammann (1981) an elliptical of flattening {\it f} will be 
called an E{\it f}, whereas an S0 galaxy of the same flattening is denoted 
S0 ({\it f}).  The physical difference between elliptical and
lenticular galaxies is that S0 galaxies contain an old disk,
whereas ellipticals do not. It often becomes difficult to 
unambiguously distinguish between these two classes of objects
when either (1) only a small fraction of the light originates
in the disk, or (2) if the disk is viewed almost pole-on. This
effect is clearly seen in the Coma cluster (Abraham \& van den Bergh 
2004) where most flattened, and hence presumably edge-on, early-type galaxies were classified as S0s, whereas those objects that
have more circular isophotes are mostly classified as ellipticals.
Nevertheless, some nearly circular [S0 (0) or S0 (1) in the 
notation of Sandage \& Tammann (1981)] galaxies are classified as 
S0 rather than E0 or E1. What are these almost circular objects 
that Sandage and Tammann (1981) classify as being of type S0 (0)?

\section{THE NATURE OF S0 (0) GALAXIES.}

   The classifications of galaxies given in {\it A Revised Shapley
-Ames Catalog of Bright Galaxies} (Sandage \& Tammann 1981) represent
the gold standard of galaxy classification because they were almost 
all based on inspection of photographic images obtained with large
reflectors, that were classified in a uniform fashion by highly 
experienced galaxy morphologists. The Shapley-Ames Catalog (in which 
$H_{o}$ = 50 km $s^{-1}$ Mpc$^{-1}$ was assumed) contains 45 ellipticals of types 
E0 + E1 which have $<M_{B}^{o}>$ = -21.19 $\pm$ 0.15 and 14 S(0) (0)
plus S0 (1) galaxies for which $<M_{B}^{o}>$ = -20.25 
$\pm$ 0.30.
The mean luminosity difference between E0 and E1 galaxies
on the one hand, and S0 (0) and  S0 (1) galaxies on the
other, is found to be 0.94 $\pm$ 0.34 mag. However, this
estimate of the mean error in the difference between the
mean luminosities makes the unwarranted assumption that
both E0 +E1 and the S0 (0) plus S0 (1) galaxies in the Shapley-Ames Catalog have Gaussian luminosity distributions. It is therefore better to use a non-paramentric test to assess the significance of the difference in the mean luminosities of nearly  circular E and S0 galaxies. Such a test is provided by the data in Table 1 which shows the number of nearly circular E and S0 galaxies that are brighter (or fainter) than $<M_{B}^{o}>$ = -20.5.  For the data in Table 1  Chi-sqared = 7.2 which, for one degree of freedom, yields an a-priori probability of only 0.7\% for the hypothesis that nearly round ellipticals and nearly circular S0 galaxies have parent populations with the same luminosity distribution. [A Kolmogorov-Smirnov test yields a probability of 11\% that these two samples were drawn from the same parent population.] In other words the present data strongly suggest that E0 + E1 galaxies are systematically more luminous than S0 (0) and S0 (1) galaxies. This means that assigning a round early-type galaxy to the E or S0 class may be regarded as a crude form of luminosity classification. Whereas the morphological luminosity classification of spiral galaxies (van den Bergh 1960abc) was based on the characteristics of their spiral arms, the distinction between E and S0 galaxies is based on the existence (or absence) of a faint amorphous envelope Sandage (1961,
p.11).

The galaxy luminosities given by Sandage \& Tammann
(1981) are on the $B_{T}$ system of the RC2 Catalog of
de Vaucouleurs et al. (1976). However, because
integrated magnitudes depend on the outer profiles
of galaxies, there is a small systematic difference
between absolute magnitudes of galaxies on the $B_{T}$
system of de Vaucouleuers et al. and those on the
$B_{26}$ system of Sandage \& Visvanathan (1978). According
to Sandage \& Tammann (1981, p.7) ($B_{26} - B_{T}$) = -0.12
mag for E's and ($B_{26} - B_{T}$) = -0.03 mag for S0
galaxies. However, these effects are seen to be an
order of magnitude smaller than the observed
systematic differences between the luminosities of
nearly spherical E and S0 galaxies. In other words
the systematic difference between the luminosity
distributions of E and S0 galaxies is too large to
be attributed to systematic differences between the
luminosity profiles of these two types of objects.
    Inspection of Fig. 22 of van den Bergh (1997)
shows that S0 galaxies are not only fainter than
E0 galaxies but they are, on average, also fainter
than galaxies of types Sa and Sb. This shows that
the luminosity differences between spiral, S0, and
E galaxies are not primarily due to the systematic
changes of galaxy luminosity along the Hubble
classification sequence.

\section{CONCLUSION.}

   Among very early-type galaxies with almost circular isophotes
some are called ellipticals [E0 + E1] and others are classified as
lenticulars [S(0) (0) and S(0) (1)]. It is shown that the objects 
that were classified as ellipticals are, on average, almost one 
magnitude more luminous than those that are called S0. This suggests 
that the dichotomy between round ellipticals and face-on S0 galaxies 
(which appear to have extended amorphous envelopes) represents a 
crude form of luminosity classification for early-type galaxies.  Physically this result confirms that rotation and ``diskiness'' are more important in the outer regions of faint early-type galaxies than they are in more luminous early type objects.  This conclusion is consistent with earlier work by Davies et al. (1983), Capaccioli et al. (1990) and by Rix, Carollo \& Freeman (1999).

\begin{deluxetable}{lcc}
\tablecaption{Luminosity versus type for round early-type galaxies in the Shapley-Ames catalog\label{tbl-1}}
\tablehead{\colhead{Types} & \colhead{$M_{B}^{o} \leq -20.50$} & \colhead{ $M_{B}^{o}> -20.50$}}

\startdata

S0 (0) + S0 (1)    &     $ 6$        &      $8$\\  
E0  + E1           &     $36$        &      $9$\\

\enddata

\end{deluxetable}

\end{document}